# Na-catalyzed rapid synthesis and characterization of intercalated graphite $CaC_6$

Akira Iyo,[a,*] Hiroshi Fujihisa,[a] Yoshito Gotoh,[a] Shigeyuki Ishida,[a] Hiroshi Eisaki,[a] Hiraku Ogino,[a] Kenji Kawashima[a,b]

[a]National Institute of Advanced Industrial Science and Technology (AIST), Tsukuba, Ibaraki 305-8568, Japan
[b]IMRA JAPAN Co., Ltd., Kariya, Aichi 448-8650, Japan

*Corresponding author: E-mail address: iyo-akira@aist.go.jp

## Abstract

In this study, we conducted experiments on $CaC_6$ for elucidating the Na-catalyzed formation mechanism and achieving rapid mass synthesis of graphite intercalation compounds (GICs). Rapidly synthesized $CaC_6$ was characterized by analysis of its crystal structure and physical properties. We found that the formation of the reaction intermediate Na-GIC ($NaC_x$, $x \approx 64$) requires a larger amount of Na than is intercalated between the graphite interlayers. The requirement for excess Na may provide insights into the mechanism of Na-catalyzed GIC formation. A Na-to-C molar mixing ratio of 1.5–2.0:6 was suitable for the efficient formation of $CaC_6$ under heat treatment at 250 °C for 2 h, and the catalytic Na remaining in the sample was demonstrably reduced to a Na:Ca ratio of approximately 3:97. The upper critical field $H_{c2}$ was enhanced approximately three times compared to those of previous reports. Based on X-ray diffraction and experimental parameter analysis, we concluded that the enhancement of $H_{c2}$ was attributed to the disordered stacking sequence in $CaC_6$, possibly because of the rapid and low-temperature formation. Physical properties derived from specific heat measurements were comparable to those of high-quality $CaC_6$, which is slowly synthesized using the molten Li-Ca alloy method. This study provides new avenues for future research and exploration in the rapid mass synthesis of GICs as practical materials, for applications such as battery electrodes and superconducting wires.

## 1. Introduction

Graphite is a two-dimensional layered material composed of stacked graphene, and graphite intercalation compounds (GICs) are formed by the insertion of atoms or molecules between these graphene layers. GICs exhibit various functions depending on the intercalated atoms or molecules [1]. Intercalation of lithium into graphite used in lithium-ion secondary batteries is its best-known application [2,3]. The discovery of superconductivity in $CaC_6$ with a transition temperature as high as 11.5 K demonstrates the further potential of GICs [4,5]. While research into the applications of GICs is actively underway [6], it is also essential to develop efficient synthetic methods for GICs. For GICs intercalated with donor-type atoms, such as alkali metals ($A_M$) and alkaline earth metals ($A_E$), effective synthetic techniques, such as vapor phase [7] and molten salts methods [8], have been proposed. Although these methods can be used to synthesize high-quality GICs, they are not necessarily suitable for the rapid mass synthesis required for feasible applications.

In recent studies, the catalytic properties of Na were found to accelerate the formation of $A_M$ (Li, Na, K) or $A_E$ (Ca, Sr, Ba) intercalated graphite [9]. Specifically, simply mixing $A_M$ and graphite (C) with Na at room temperature (RT, ~25 °C) yielded graphite intercalation compounds ($A_M$-GICs) with controlled stage structures. $A_E$-GICs ($A_E C_6$) were formed by heating a mixture of $A_E$ C, and Na at 250 °C for only 2–4 h. High-stage Na-GIC ($NaC_x$, $x \approx 64$), formed by mixing Na and C, was proposed to be a reaction intermediate of the catalyst that significantly reduces the activation energy for intercalation.

The Na-catalyzed method paves the way for efficient and rapid mass production of $A_M(A_E)$-GICs, thereby promoting research and development for practical applications of GICs, including their use as active electrode materials for rechargeable batteries. In addition, the catalytic function of Na lowers the intercalation barrier to the graphite interlayers and thus holds the potential for enhancing the performance of $A_M(A_E)$ ion batteries. However, investigations into the Na-catalyzed synthesis of GICs are still in their early stages, and various issues remain to be addressed. Elucidation of the mechanism of Na-catalyzed GIC formation, optimization of GIC synthesis conditions, and evaluation of GICs rapidly synthesized by the Na-catalyzed method are fundamental issues to be studied.

In this study, $CaC_6$ is chosen as the GIC to be investigated to address the above issues. $CaC_6$ is one of the most challenging GICs to synthesize [10] and is conventionally synthesized slowly (typically over 10 days) at 350 °C using the Li-Ca melting alloy method with highly oriented pyrolytic graphite (HOPG) as the host material [5]. Thus, $CaC_6$ is suitable for comparing the properties of samples from contrasting synthesis methods. Furthermore, the superconducting properties of high-quality $CaC_6$ have been studied in detail, providing an appropriate benchmark for comparisons. Notably, $CaC_6$ itself is considered an attractive material with potential for application as a negative electrode active material in Ca ion batteries [11].

We first investigated the effect of the Na-to-material mixing ratio on the formation of the reaction intermediate $NaC_x$, which plays a key role in Na-catalyzed GIC formation. Consequently, this experiment provided important insights into $NaC_x$ formation. Next, we performed experiments to obtain a guideline of the Na-to-material mixing ratio for the efficient production of $CaC_6$ for practical materials. The samples synthesized by the Na-catalyzed method are inevitably a mixture of GICs and Na. To accurately evaluate the properties of Na-catalyzed GICs, we developed a method for reducing the residual Na in the synthesized samples. The reduction of the residual Na is essential for producing GICs as practical materials because Na can have a negative impact on performance in certain applications.

Finally, the crystal structure and superconducting properties of the rapidly synthesized $CaC_6$ were evaluated by powder X-ray diffraction (XRD) analysis, upper critical field measurements, and specific heat measurements. The properties were compared to those of the slowly synthesized HOPG-based $CaC_6$ to examine whether there are differences due to the synthesis method. This study provides insights into the mechanism of Na-catalyzed GIC formation, rapid mass synthesis of GICs, potential applications, and avenues for future research on this new synthetic method.

## 2. Material and methods

*2.1 Sample preparation to study the effect of the Na mixing ratio*

Powdered graphite (Furuuchi Chemical, 99.99%, –200 mesh) was used as the intercalation host. A Ca lump (Furuuchi Chemical, 99.5%) was filed to powder (~50 μm) before weighing. A lump of soft Na metal (Furuuchi Chemical, 99.9% purity) was cut and weighed.

To investigate the effect of the Na-to-material mixing ratio on the formation of $NaC_x$ or $CaC_6$, the materials were weighed to obtain a Ca:C:Na ratio of 1:6:$y$ ($y$ = 0, 0.5, 0.75, 1.0, 1.5, 2.0, and 2.5), where $y$ represents the molar mixing ratio of Na to $CaC_6$. The reagents (~0.1 g in total) were placed in a zirconia mortar (**Fig. 1a**) and mixed using a pestle at RT for ~15 min (**Fig. 1b**). Weighing and mixing of reagents were performed in a glovebox (GB) filled with Ar gas. The mixed sample was placed in a pellet-forming die with an inner diameter of 6.7 mm, which functioned as a reaction vessel, and heated on a hot plate at 250 °C for 2 h (**Fig. 1c**) in GB.

*2.2 Sample preparation for the measurement of physical properties*

The materials weighed to obtain a Ca:C:Na ratio of 1:6:2 ($y$ = 2) were mixed at RT for ~15 min. The sample mixture was heated at 250 °C for a total of 4 h in the pellet-forming die with intermediate mixing [9]. The heat treatment yielded a mixture of $CaC_6$ and Na. The Na in the sample was reduced using the following two-step process to minimize the effect of residual Na on its physical properties.

In the first step, the sample was heated to 150 °C (above the melting point of Na) in the pellet-forming die on a hot plate (**Fig. 1c**). Then, the pellet-forming die was moved to a hand press, and a pressure of 300 MPa was applied to the pushing rod to squeeze out the melted Na from the sample (**Fig. 1d**). This treatment lowered the Na/Ca ratio from 2:1 ($y$ = 2) to 1:3. Notably, the squeezed-out Na can be reused because it does not contain Ca.

In the second step, the sample, in pellet form, was sealed in a quartz tube under vacuum (~ 0.05 torr), heated to 300 °C over 30 min and held for 18 h, while one side of the quartz tube was kept at RT. The evaporated Na from the pellet reacted with the quartz tube and was deposited onto the part of the inner wall kept at RT (**Fig. 1e**). This treatment further reduced the Na/Ca ratio to ~3:97. A photograph of a sintered pellet obtained using this process is shown in **Fig. 1f**. The pellets have a light golden color and a density of 2.20 g cm$^{-3}$ (87% of the theoretical density). The pellet can easily be ground into powder to serve as materials for batteries, superconducting wires, and other applications (**Fig. 1g**).

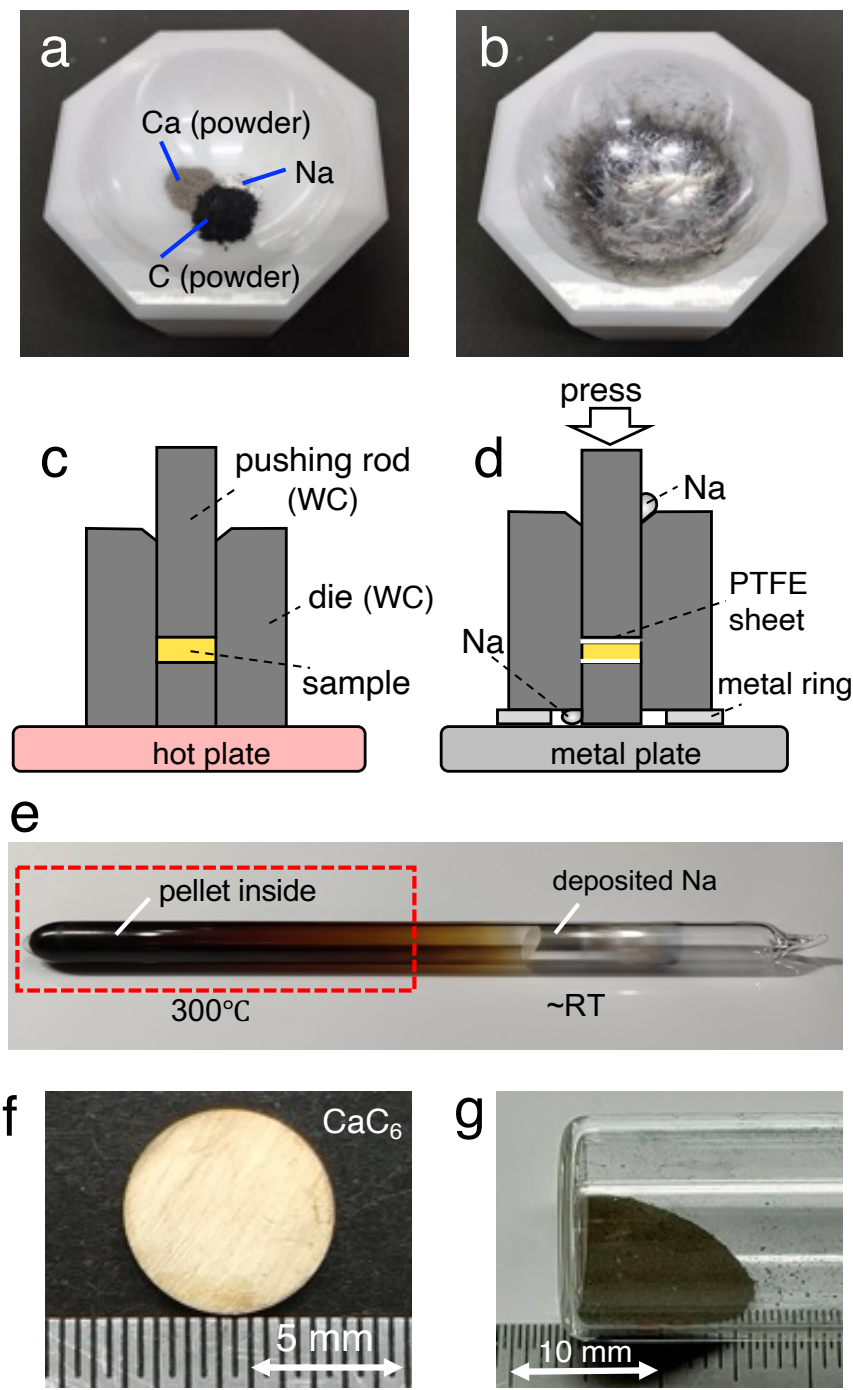

Fig. 1 (a) Weighed reagents (Ca powder, C powder, and Na lump) in a zirconia mortar. (b) Sample obtained by mixing the reagents with a pestle at RT for ~15 min. (c) Heat treatment of the sample in a pellet-forming die made of tungsten carbide (WC) on a hot plate. (d) Pressing to squeeze out Na from a sample. Before pressurization, the pellet-forming die and sample are heated at 150 °C. The sample is sandwiched between polytetrafluoroethylene (PTFE) sheets to prevent adhesion to the pushing rod. (e) Quartz tube with sample inside, heat-treated to further reduce Na in the sample. One side of the quartz tube is held at RT. (f) Image of the sintered $CaC_6$ pellet (6.7 mm diameter, ~1.0 mm thick). (g) Image of the $CaC_6$ powder (~0.2 g) in a bottle, obtained by grinding the pellet.

*2.3 Measurements*

The samples were evaluated using XRD with Cu$K_\alpha$ radiation (Rigaku Ultima IV). XRD measurements were performed using an airtight attachment to prevent exposure of the samples to air. The elemental composition of the samples was analyzed using an electron microscope (Hitachi High-Technologies, TM-3000) equipped with an energy-dispersive X-ray spectrometer (Oxford, SwiftED3000).

The temperature ($T$) and magnetic field ($H$) dependence of the magnetization ($M$) was measured using a magnetic property measurement system (Quantum Design, MPMS-XL7). The $T$ dependence of the electrical resistivity ($\rho$) and specific heat ($C_p$) was measured using a physical property measuring system (Quantum Design, PPMS). Electrodes for $\rho$ measurement using the four-terminal method were attached to the sample in the GB using a silver paste (Dupont 4922N).

## 3. Results and discussion

*3.1 Effect of the Na mixing ratio on $NaC_x$ formation*

The XRD patterns of the samples mixed at RT (as-mixed samples) as a function of $y$ are shown in **Fig. 2a**. The diffraction intensities were normalized to the peak intensity ($I_p$) of Ca, which did not react by mixing at RT. Peaks indicating the formation of the reaction intermediate $NaC_x$ were observed, along with the peaks of the starting materials Ca, C, and Na. **Fig. 2b** shows the dependence of normalized $I_p$ on the parameter $y$ for Ca, C, Na, and $NaC_x$.

For samples with $y = 0$ and 0.25, only Ca and C peaks appeared in the XRD patterns, and the $NaC_x$ peak began to appear at $y = 0.5$. The normalized $I_p$ of $NaC_x$ increased with $y$ and remained nearly constant at $y \geq 1.0$ owing to the completion of the C and Na reaction. As reported previously [9], the diffraction peak of $NaC_x$ at $2\theta \approx 25.5°$ has a $d$-value of ~3.49 Å, which is close to the expected $d$-value of 3.50 Å for $NaC_x$ in stage 8 ($x = 64$) [12]. However, the broadening of the diffraction peak compared to that of the host graphite is attributed to non-uniform staging (around stage 8). At $y = 0.5$, where $NaC_x$ begins to form, $I_p$ values other than that of Ca are extremely low. This means that the staging is highly irregular owing to the ongoing intercalation of Na into the graphite interlayers.

Based on quantitative considerations, a Na:C ratio of 1:60 ($y = 0.1$) should have been sufficient for $NaC_{64}$ formation. However, even at a higher ratio of Na:C = 1:24 ($y = 0.25$), neither $NaC_x$ nor Na peaks appeared. This result indicates that for $NaC_x$ to form, excess Na must be mixed with C (greater than the amount intended for intercalation in the graphite interlayers). In addition, the absence of Na peaks up to $y = 0.5$ indicates that the Na not inserted in the graphite interlayer is not in the form of elemental Na. Similar to the binding of Li at the graphite edge proposed in Li-GICs [13], the binding of Na at the graphite edge possibly explains the necessity for excess Na and the presence of non-elemental Na; further, this may be related to the formation of $NaC_x$ that is evaluated as unstable in theoretical calculations [14-16].

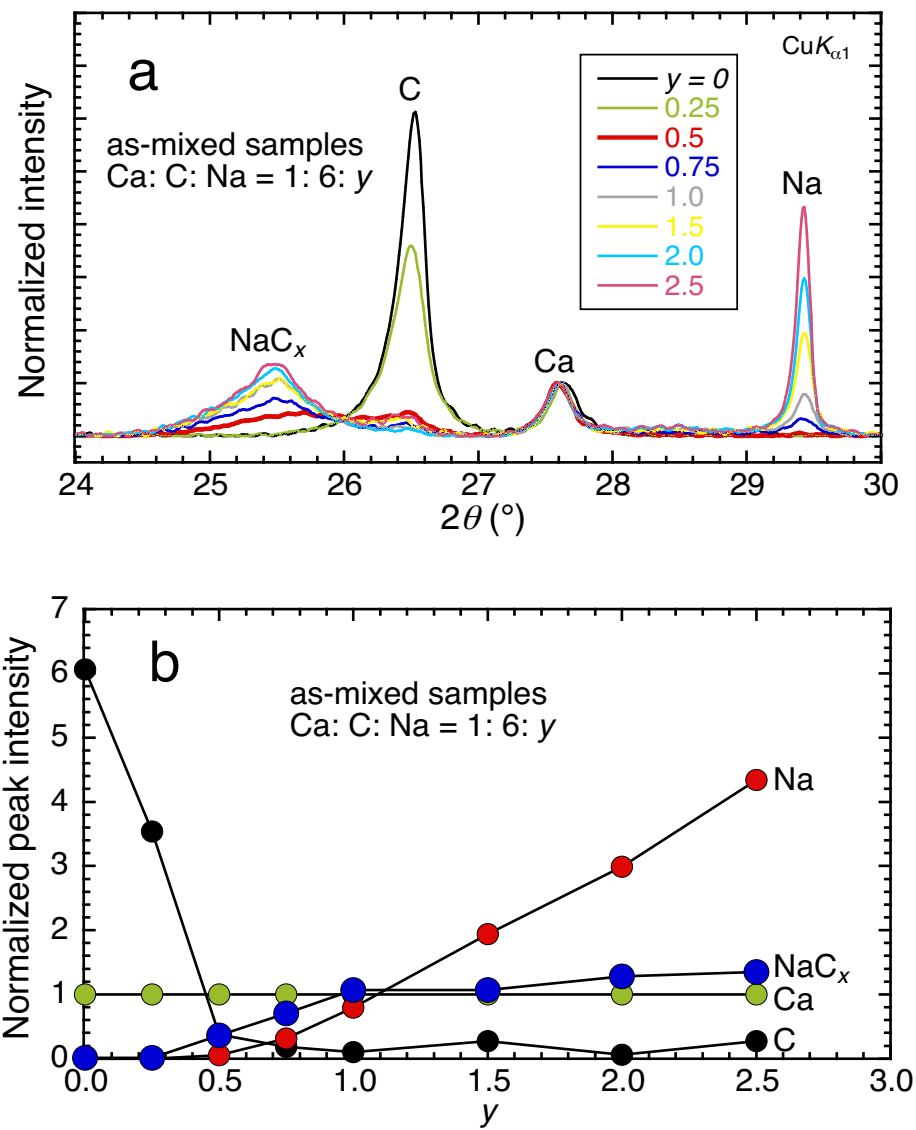

Fig. 2 (a) XRD pattern of a sample mixed at RT as a function of $y$. The background and $CuK_{\alpha2}$ diffraction components were eliminated from the data via software processing. The diffraction intensities are normalized by the peak intensity of Ca, which did not react at RT. (b) $y$ dependence of the normalized $I_p$ of Ca, C, Na, and $NaC_x$.

*3.2 Effect of the Na mixing ratio on $CaC_6$ formation*

The XRD patterns of the samples obtained by heating the above as-mixed samples at 250 °C for 2 h and the $y$ dependence of $I_p$ on Ca, C, Na, and $CaC_6$ are shown in **Figs. 3a and b**, respectively. For comparison of the diffraction intensities, the samples were placed on sample holders and pressed hard, resulting in strongly oriented $CaC_6$ diffraction patterns.

In the samples with $y = 0$ and 0.25, only a trace amount of $CaC_6$ was formed. The $I_p$ of $CaC_6$ increased with $y$ for $y \geq 0.5$ and showed a broad maximum at $y = 2.0$. Under the heat-treatment conditions of 250 °C for 2 h, the optimal value of $y$ for efficient $CaC_6$ formation was estimated to be 1.5–2.0. The elemental Na not contributing to $NaC_x$ formation, detected as diffraction peaks, probably plays a role in promoting Ca diffusion as a solvent. Therefore, it should be noted that adding too much Na may slow down the GIC formation. Further, the optimal $y$ would naturally vary depending on the heat-treatment temperature and duration. For example, if the reaction duration is lengthened, less $y$ should be required to complete the GIC formation.

The $CaC_6$ peak for $y = 0.5$ was much smaller than that for $y \approx 2.0$, although the C and Ca peaks were almost absent. This suggests non-uniform staging due to the high-stage $NaC_x$ being in the process of reacting with Ca toward first-stage $CaC_6$. The microscopic formation process of GICs, i.e., how it is converted from $NaC_x$ to $CaC_6$, is not clear from this study. This is a critical issue for the Na-catalyzed GIC synthesis and must be studied in detail both experimentally and theoretically in the future.

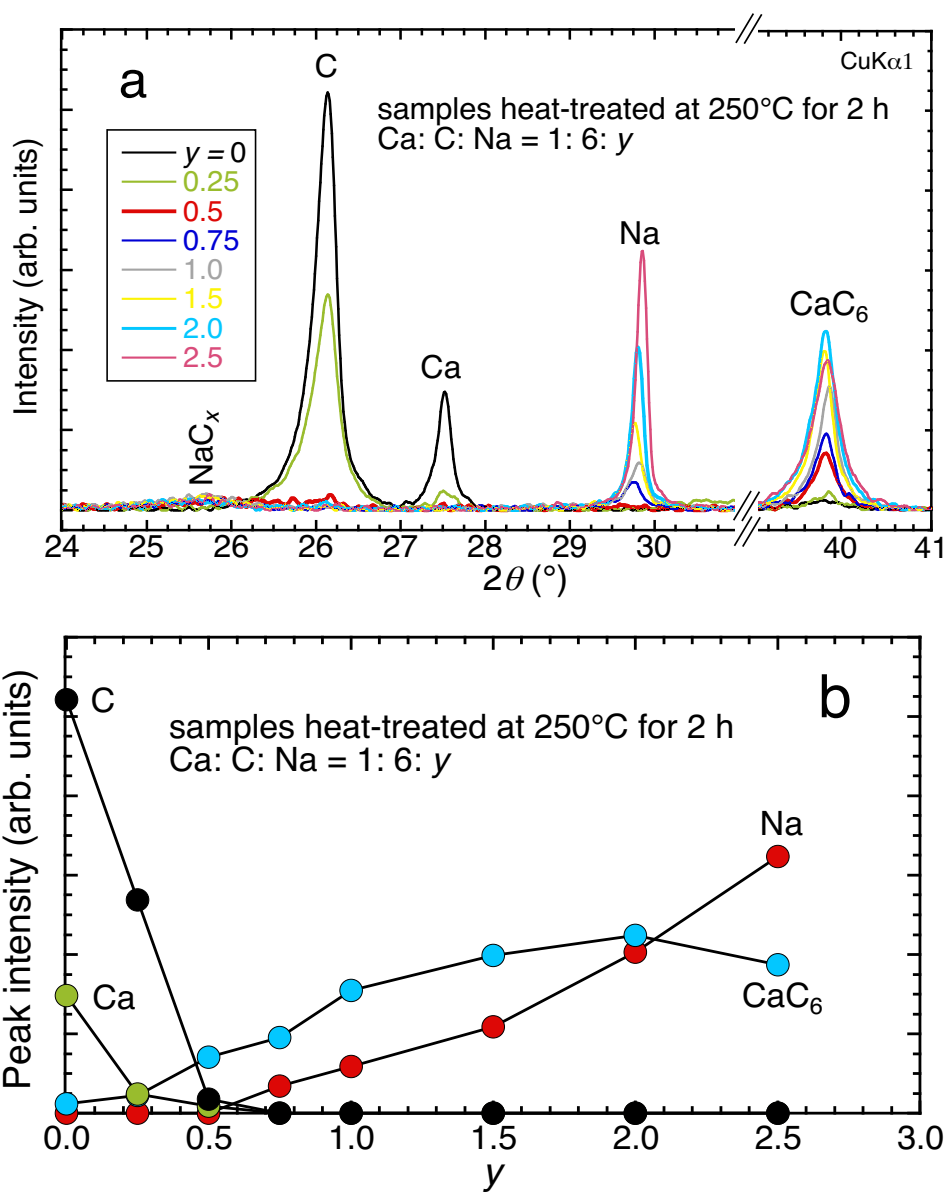

Fig. 3 XRD patterns of the samples obtained by heating the samples for $y = 0, 0.5, 0.75, 1.0, 1.5, 2.0, 2.5$ at 250 °C for 2 h. The background and $CuK_{\alpha2}$ diffraction components were eliminated from the data via software processing. (b) $y$ dependence of the $I_p$ of Ca, C, Na, and $CaC_6$. $I_p$ of $NaC_x$ are not included because they exhibit very low intensity for all $y$.

*3.3 Sample evaluation for measuring superconducting properties*

The XRD pattern of the powder sample obtained by grinding the pellet used for measuring the superconducting property is shown by the red dots in **Fig. 4a**, and the Miller indices are labeled on diffraction peaks, assuming the *R-3m* structure [17]. The powder XRD patterns were analyzed using Materials Studio Reflex [18] version 2022HF1. When the powder had no preferred orientation, the intensity ratio of the peaks at 006 to 113 ($R_{006/113}$) was ~0.56. However, the powdered sample was preferentially oriented in the 00*l* direction, resulting in extremely weak non-00*l* reflections ($R_{006/113}$ ~5) [9]. We carefully set the powder on the measurement plate to minimize the preferred orientation, thereby reducing $R_{006/113}$ to ~1.1. The parameter $R_0$ of the March-Dollase function [19] estimated from $R_{006/113}$ was $R_0 = 0.84$. The blue profile in **Fig. 4a** shows the simulated pattern of the *R-3m* structure at $R_0 = 0.84$. The lattice constants *a* and *c* (hexagonal setting) determined from the diffraction pattern were 4.330(1) and 13.593(1) Å, respectively, almost identical to those of HOPG-based $CaC_6$ [17].

Graphite stacking in the stage-1 GICs represents only A-stacking with hexagonal holes aligned in the *c*-axis direction [17]. In contrast, Ca atoms have three locations: *α*, *β*, and *γ* [17]. If Ca atoms entered these sites with an equal probability of 1/3, the peaks around 24° to 29° would disappear. The presence of peaks in this region indicates that the arrangement of Ca atoms is not entirely random but has some periodicity in the *c*-axis direction. For the *R-3m* structure, 101 and 012 reflections appear in this region. Comparing the blue simulated profile in **Fig. 4a** with the red observed pattern, the peak width in the region is broader than those of the 00*l* peaks, indicating that the *αβγ* periodicity of the Ca atoms is not definite and contains stacking faults. Stacking, such as *αβα*, for example, would also exist locally. Comparing the time and temperature required for sample synthesis, the molten Li-Ca alloy method requires 10 days at 350 °C [5], whereas our method takes less than 24 h at below 300 °C, including the Na reduction heat treatment. This difference in synthesis conditions is possibly one of the causes leading to the difference in the regularity of the stacking sequence.

**Fig. 4b** shows the *T* dependence of $4\pi M/H$ for a rectangular sample cut from the pellet. Demagnetization correction was performed for *M* according to the sample geometry. As shown in the inset, the onset temperature of the transition ($T_c^{on}$) of the Na-catalyzed $CaC_6$ was 11.0 K, which is 0.5 K lower than that of the HOPG-based $CaC_6$ [4,5]. The above-mentioned disordered stacking sequence is a possible cause of the slightly lower $T_c$. The midpoint temperature of the transition ($T_c^{mid}$) determined from the field-cooling (FC) magnetization was 10.5 K. The shielding volume fraction of superconductivity exhibited by zero-FC magnetization was ~100% ($4\pi M/H = -1$), indicating that the pellet was well-sintered.

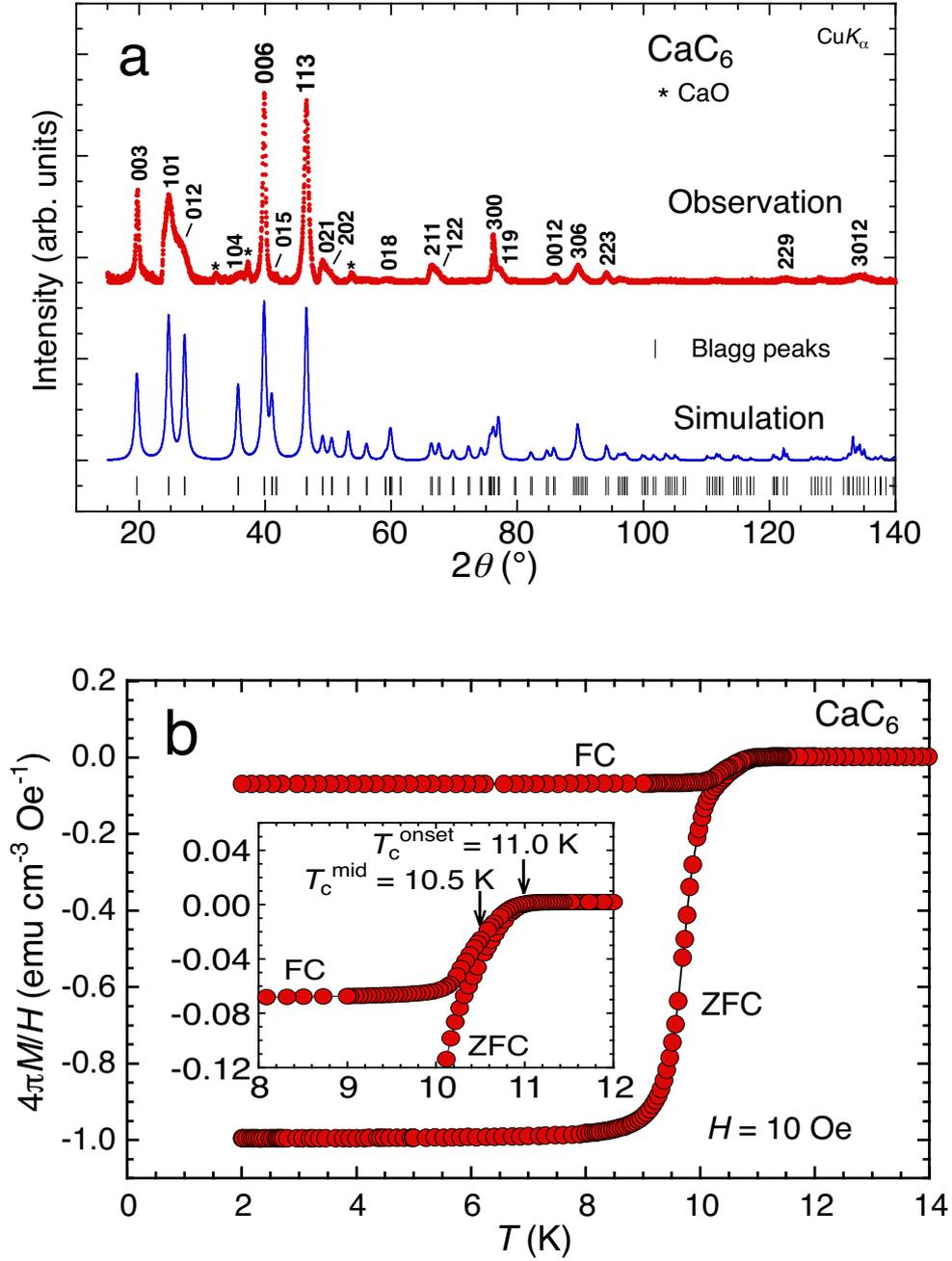

Fig. 4 (a) Observed XRD pattern of CaC$_6$ (red dots) and simulated XRD pattern for the *R*-3*m* (blue profile). Numbers labeled on the diffraction peaks represent Miller indices assuming the *R*-3*m* structure [17]. Asterisk marks are diffraction peaks from impurity CaO. (b) *T*-dependent $4\pi M/H$ of the rectangular polycrystalline sample of CaC$_6$. The inset is a magnified view near the superconducting transition.

*3.4 Upper critical field measurement*

    **Fig. 5a** shows the XRD pattern measured for the pellet surface. As is clear from the comparison with **Fig. 4a**, the *c*-axes of the CaC$_6$ crystals in the pellet are highly oriented perpendicular to the pellet surface. As illustrated in the inset of **Fig. 5b**, $\rho$ was measured in the configuration in which four terminals were attached to a cross section of a rectangular sample cut from the pellet, and magnetic fields were applied parallel to the pellet surface, viz., *H*//*ab*.

Although the CaC$_6$ crystals in the sample are not perfectly aligned, this configuration provides a nearly accurate $H_{c2}//ab$ measurement, as $H_{c2}//ab$ is approximately 3–5 times larger than $H_{c2}//c$ [20-23].

The main panel of **Fig. 5b** shows the $T$-dependent $\rho$ of the CaC$_6$ sample in $H = 0$. In sharp contrast to the behavior of HOPG-based CaC$_6$ [23,24], $\rho$ exhibited a linear temperature dependence, except in the low-temperature region (< ~50 K). The resistivity at 300 K ($\rho_{300K}$) and residual resistivity ($\rho_{res}$) were 132 and 59.2 μΩ cm, respectively. Owing to its polycrystalline nature, the CaC$_6$ pellet exhibited a significantly small residual resistivity ratio ($RRR = \rho_{300K}/\rho_{res} = 2.2$) compared to the HOPG-based CaC$_6$ [24]. The inset of **Fig. 5b** shows the $T$-dependent $\rho$ below $T_c$ in $H$ up to 30 kOe. As indicated in the inset, $T_c^{on}$ at $H = 0$ is 11.0 K, which agrees with the measured value via magnetization.

**Fig. 5c** shows the $T$ dependence of upper critical field $H_{c2}(T)$ defined at the midpoint $T$ where $\rho$ decreases to 50% of that in the normal state, together with those reported for the HOPG-based CaC$_6$ [20-23]. $H_{c2}(T)$ of the present study showed a $T$-linear dependence characteristic of CaC$_6$ and others depicted in **Fig. 5c**. The linear extrapolation of $H_{c2}(T)$ yielded an $H_{c2}(0)$ of 38.1 kOe as indicated by the dotted line, and the Ginzburg–Landau coherence length at 0 K ($\xi$) was calculated to be 92.9 Å using $H_{c2}(0) = \Phi_0/(2\pi\xi^2)$, where $\Phi_0$ is the magnetic flux quantum. The $H_{c2}(0)$ value in this study was approximately three times larger than those measured for the HOPG-based CaC$_6$. Shortening of $\xi$ due to the stacking sequence disorder suggested by powder XRD analysis can be attributed to the enhancement of $H_{c2}$, similarly to the enhanced $H_{c2}$ by the introduction of crystal disorder in Nb$_3$Sn [25]. This point will be discussed below, based on the physical parameters obtained by the specific heat measurement.

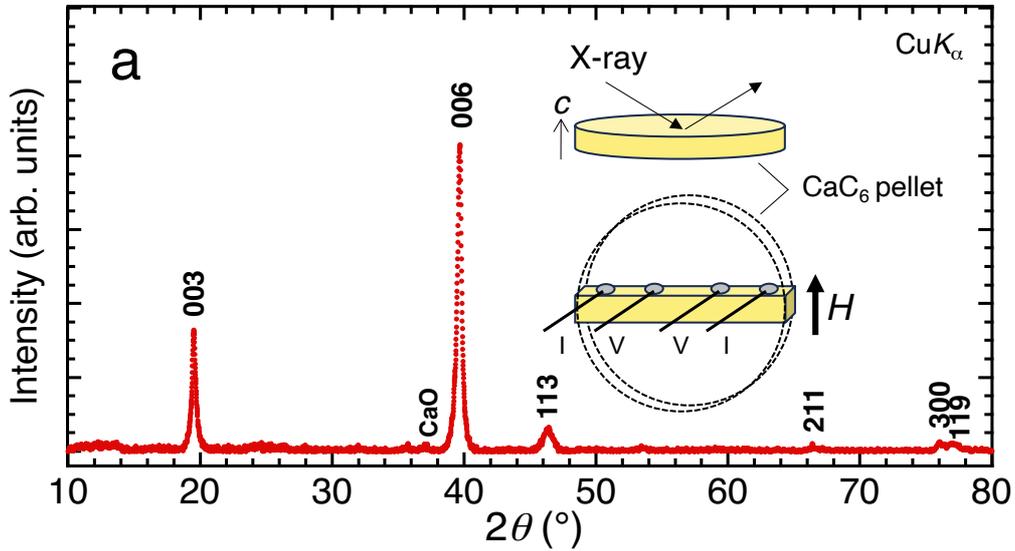

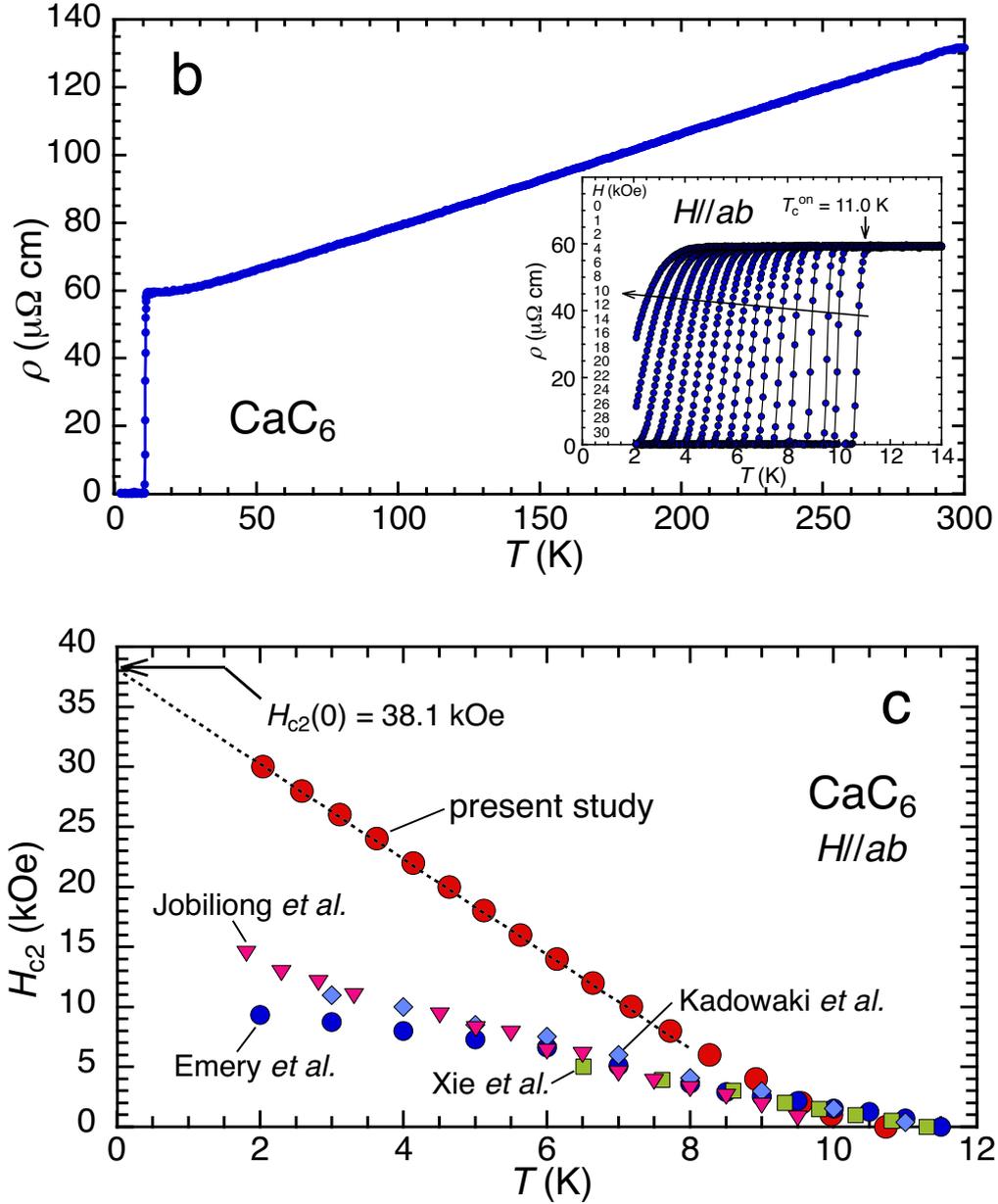

Fig. 5 (a) XRD pattern of the pellet surface. The inset shows the configurations of the XRD measurement of the pellet, the four terminals of the rectangular sample cut from the pellet, and the direction of the applied magnetic field. (b) $T$-dependent $\rho$ of the $CaC_6$ pellet in $H = 0$. The inset shows the $T$-dependent $\rho$ below the approximate $T_c$ in $H$. (c) $T$-dependent $H_{c2}$ determined from the midpoint $T_c$. $H_{c2}$ reproduced from previous studies [20-23] are those measured by applying a magnetic field parallel to the $ab$ plane ($H_{c2}//ab$) of HOPG-based $CaC_6$.

*3.5 Specific heat measurement*

**Fig. 6** illustrates the dependence on $T^2$ of $C_p/T$ for the $CaC_6$ sample at $H = 0$ and 50 kOe, sufficient to suppress superconductivity. The large anomaly in $C_p/T$ at $H = 0$ indicates the bulk superconducting nature of the samples. As indicated by the dashed line in **Fig. 6**, a distinct deviation from Debye's $T^3$ law was observed. Such deviations are commonly observed in $CaC_6$ [26] and $A_M$-GICs [27-29].

$C_p$ at $H$ = 50 kOe below $T$ < 12 K was fitted using $C_p(T) = C_{el}(T) + C_{ph}(T)$, where $C_{el}(T)$ (= $\gamma_n T$) and $C_{ph}(T)$ (= $\beta T^3 + \delta T^5$) are the electron and lattice contributions to the specific heat, respectively. The fitting yielded $\gamma_n$ (Sommerfeld coefficient) = 6.68 mJ mol$^{-1}$ K$^{-2}$, $\beta$ = 92.9 μJ mol$^{-1}$ K$^{-4}$, and $\delta$ = 0.335 μJ mol$^{-1}$ K$^{-6}$. Debye temperature ($\Theta_D$) was calculated to be 527 K using $\beta = 12N\pi^4 R\Theta_D^{-3}/5$ where $N$ (= 7) is the number of atoms in a formula unit cell and $R$ (= 8.314 J mol$^{-1}$ K$^{-1}$) is the gas constant. The $\Theta_D$ was slightly lower than the 593 K, measured for the HOPG-based $CaC_6$ [26], possibly due to the stacking sequence disorder, which may explain the slightly low $T_c$ of the Na-catalyzed $CaC_6$.

The inset shows the $T$ dependence of $C_{el}/T$, where $C_{el}$ was obtained by subtracting $C_{ph}$ from the total $C_p$. The onset and midpoint $T_c$ determined from the specific heat jump were 11.0 and 10.5 K, respectively, in agreement with those obtained from the $M$ measurement. In the following analysis, 10.5 K was employed as $T_c$. Non-zero $C_{el}/T$ at a 0 K limit indicates the presence of small residual $\gamma_{res}$ (= 0.720 mJ mol$^{-1}$ K$^{-2}$). The $\gamma_{res}$ can be attributed to residual Na and non-superconducting CaO, whose presence was confirmed by the composition and XRD measurements, respectively. The Sommerfeld coefficient of the superconducting part of the sample $\gamma_s$ (= $\gamma_n$ - $\gamma_{res}$) is 5.96 mJ mol$^{-1}$ K$^{-2}$, which agrees well with the reported value of 6.01 mJ mol$^{-1}$ K$^{-2}$ [26].

The $T$ dependence of $C_{el}/T$ fits well with the predictions of the Bardeen–Cooper–Schrieffer (BCS) theory, as indicated by the solid curve. Using $\Theta_D$ = 527 K, $T_c$ = 10.5 K, and Coulomb pseudopotential $\mu^*$ = 0.15 (0.10) in the MacMillan equation $T_c = (\Theta_D/1.45) \exp[-1.04 (1 + \lambda_{e-p}) / (\lambda_{e-p} - (1 + 0.62\lambda_{e-p}) \mu^*)]$, the electron-phonon coupling constant $\lambda_{e-p}$ = 0.72 (0.61) was obtained. Inputting $\gamma_s$ and $\lambda_{e-p}$ = 0.72 (0.61) to $N(E_F) = 3\gamma_s/[\pi^2 k_B^2 (1 + \lambda_{e-p})]$, where $k_B$ is the Boltzmann constant, the theoretically derived density of states at the Fermi energy $N(E_F)$ was calculated to be 1.47 (1.57) eV$^{-1}$ f. u.$^{-1}$, which agrees well with that theoretically derived in previous studies, ~1.50 eV$^{-1}$ f. u.$^{-1}$ [30,31].

**Table 1** summarizes the physical parameters of $CaC_6$ determined in this study and those reported for HOPG-based $CaC_6$. Although there are some differences, such as the smaller $RRR$ and larger $H_{c2}(0)$, owing to the polycrystalline nature and possible stacking defects in our $CaC_6$ sample, the other physical properties are comparable to those of high-quality HOPG-based $CaC_6$.

Finally, the origin of the largely enhanced $H_{c2}(0)$ is discussed in terms of the disorder effect. In a dirty superconductor, the effective coherence length is influenced by the mean free path $l$ through the relation $1/\xi = 1/\xi_{BCS} + 1/l$, where $\xi_{BCS}$ is the BCS coherence length. This indicates that the introduction of disorder, which causes the decrease of $l$, results in the shortening of $\xi$, i.e., the increase of $H_{c2}(0)$. Assuming a spherical Fermi surface, $l$ can be represented as $l = 3\pi^2\hbar^3/e^2 r_{res} m^{*2} v_F^2$, where $\hbar$ is Planck's constant divided by $2\pi$, $e$ is elementary charge, $m^*$ is effective mass, and $v_F$ is Fermi velocity. Here, the Sommerfeld coefficient $\gamma_s$ can be used to estimate $v_F$, which is related to $N(E_F)$ by $v_F = \pi^2\hbar^3 N(E_F)/m^{*2} V_{f.u.} = 3\hbar^3 \gamma_s/k_B^2 m^{*2} V_{f.u.}$, where $V_{f.u.}$ is the volume per formula unit (221 Å$^3$). Then, using $\gamma_s$ = 5.96 mJ mol$^{-1}$ K$^{-2}$ and $m^* = (1 + \lambda_{e-p})m_e = 1.72 m_e$ ($m_e$ is the free electron mass), $v_F$ is estimated to be 3.4 × 10$^7$ cm s$^{-1}$. The estimated $v_F$ is in good agreement with the average $v_F$ reported by the angle-resolved photoemission spectroscopy study (2.5 eV Å = 3.8 × 10$^7$ cm s$^{-1}$) [32]. Then, the observed $r_{res}$ (= 59.2 mΩ cm) yields $l$ = 82 Å, which is comparable with the value of $\xi$ (= 92.9 Å) obtained from $H_{c2}$, indicating the significant influence of $l$ on $\xi$. Indeed, the BCS coherence length is estimated to be $\xi_{BCS} = \hbar v_F/\pi \Delta_0 = \hbar v_F/1.76\pi k_B T_c = 440$ Å using the BCS relation $\Delta_0 = 1.76 k_B T_c$, where $\Delta_0$ is superconducting gap size. The relation $l < \xi_{BCS}$ further supports that $l$ dominates $\xi$, and the value of $\xi = 69$ Å calculated using $1/\xi = 1/\xi_{BCS} + 1/l$ agrees with the experimentally determined value. Thus, we concluded that the stacking sequence disorder in the present Na-catalyzed $CaC_6$ causes $l$ to be shorter than $\xi_{BCS}$, giving rise to a $H_{c2}$ larger than those of HOPG-based $CaC_6$.

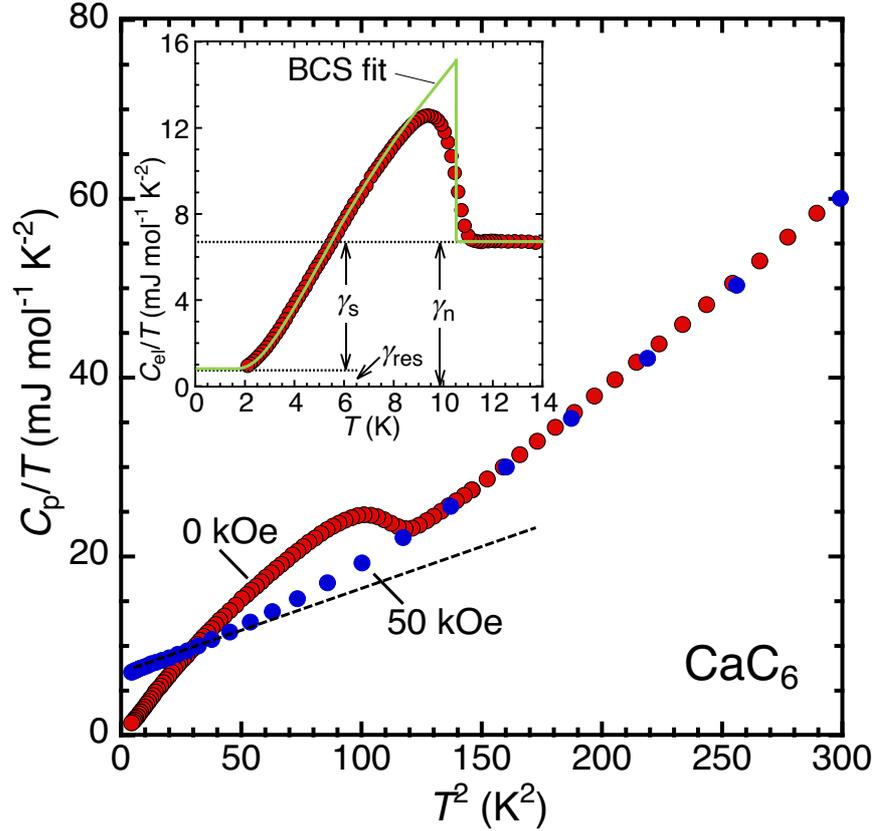

Fig. 6 $T^2$ dependence of $C_p/T$ in $H = 0$ and 50 kOe. The dashed line is a guide showing a deviation from the Debye $T^3$ rule. The inset shows $T$ dependence of $C_{el}/T$. The solid curve is fitted based on a prediction of BCS theory.

Table 1 Physical parameters determined for the Na-catalyzed $CaC_6$ and those reported for HOPG-based $CaC_6$.

| Parameters | Na-catalyzed $CaC_6$ | HOPG-based $CaC_6$ | References |
|---|---|---|---|
| $a$ (Å) | 4.330(1) | 4.333(2) | 17 |
| $c$ (Å) | 13.593(1) | 13.572(2) | 17 |
| $\rho_{300K}$ (μΩ cm) | 132 | 46 | 24 |
| $RRR$ | 2.2 | 58 | 24 |
| $T_c$ (K) | 11.0 ($T_c^{mid}$ = 10.5) | 11.5 | 4,5 |
| $H_{c2}(0)$ (kOe) ($H//ab$) | 38.1 | 12–17[a] | 20-23 |
| $\xi$ (Å) (at 0 K) | 92.9 | 139–166[a] | 20-23 |
| $\gamma_s$ (mJ mol$^{-1}$ K$^{-2}$) | 5.96 | 6.01 | 26 |
| $\beta$ (μJ mol$^{-1}$ K$^{-4}$) | 92.9 | 65.1 | 26 |
| $\Theta_D$ (K) | 527 | 593 | 26 |
| $2\Delta/k_B T_c$ | ~3.52 (BCS) | 3.55 | 26 |
| $\lambda_{e-p}$ | 0.72 | 0.71[b] | 26 |

[a] Estimated using the data in the studies [20-23].

[b] Calculated using the MacMillan equation (with $\mu^* = 0.15$) using the data ($T_c = 11.4$ K, $\Theta_D = 593$ K) in the study [26].

## 4. Conclusions

The Na-catalyzed method is a promising synthetic method for $A_M(A_E)$-GICs as practical materials. This study focused on $CaC_6$ to establish an efficient synthetic method and characterize the synthesized samples. XRD patterns of samples containing systematically varied amounts of Na indicated the presence of Na that was present neither in the graphite interlayers nor in the elemental state. A possible explanation is that Na bonded to the edge of the graphite. The role of such Na may provide insights into Na-catalyzed GIC formation and should be studied in detail with theoretical support in the future. The optimal $y$ for the efficient synthesis of $CaC_6$ was estimated to be 1.5–2 under the conditions of 250 °C for 2 h. The residual Na content in the sample was demonstrably reduced to a Na/Ca ratio of ~3:97. While the conditions of GIC synthesis and Na reduction would depend on the intended $A_M$ and $A_E$ and need to be adjusted individually, this study will provide guidance in determining those conditions. Based on the analysis and discussion of the XRD patterns and the significant enhancement of $H_{c2}$, we concluded that there is a stacking sequence disorder in the Na-catalyzed $CaC_6$ due to its rapid and low-temperature formation. This disorder affected a possible slight decrease in $T_c$ ($\Theta_D$) compared to that of HOPG-based $CaC_6$, and enhanced its practical performance as a superconducting material. Although many issues remain to be clarified by more in-depth experiments, this study is the first step in future research and development for practical materials using the Na-catalyzed method.

Acknowledgements

This study was partly conducted in collaboration with IMRA JAPAN Co., Ltd. and the National Institute of Advanced Industrial Science and Technology (AIST). This study was partly supported by the Japan Society for the Promotion of Science (JSPS) Grants-in-Aid for Scientific Research (KAKENHI) (No. 22K04193 and JP19H05823).

Declaration of competing interest

The authors declare no competing interests.

Author contributions

**A. Iyo**: Conceptualization, Methodology, Investigation, Formal analysis, Writing – Original Draft, Writing – Review & Editing. **H. Fujihisa and Y. Gotoh**: Formal analysis, Validation, Writing – Review & Editing. **S. Ishida, H. Eisaki, H. Ogino, and K. Kawashima**: Formal analysis, Writing – Review & Editing.

Data statement

The data that support the findings of this study are available from the corresponding author upon reasonable request.